\documentclass[11pt,a4paper]{article}
\usepackage{graphicx}
\usepackage{amsmath}
\usepackage{amsfonts}
\usepackage{xcolor}
\usepackage{cite}
\usepackage{authblk}
\usepackage{listings}
\usepackage{amsfonts}
\usepackage{hyperref}
\hypersetup{
     colorlinks   = true,
     citecolor    = blue,
     urlcolor = blue}
\bibliography{plain}
\title{Modeling Tripartite Entanglement in Quantum Protocols using Evolving Entangled Hypergraphs}
\author[1]{Linda Anticoli \thanks{E-mail: linda.anticoli@uniud.it}}
\author[2]{Masoud Gharahi Ghahi \thanks{E-mail: masoud.gharahi@gmail.com}}
\affil[1]{\small Department of Mathematics, Computer Science and Physics, University of Udine, Via delle Scienze 208, Udine, Italy}
\affil[2]{\small School of Advanced Studies, University of Camerino, 62032 Camerino, Italy}

\begin{document}
\maketitle

\begin{abstract}
The notion of \emph{entanglement} is the most well-known nonclassical correlation in quantum mechanics and a fundamental resource in quantum information and computation. This correlation, which is displayed by certain classes of quantum states, is of utmost importance when dealing with protocols such as quantum teleportation, quantum cryptography, and quantum key distribution. In this paper, we exploit a classification of tripartite entanglement by introducing the concepts of \emph{entangled hypergraph} and \emph{evolving entangled hypergraph} as data structures suitable to model quantum protocols which use entanglement. Finally, we present a few examples to provide applications of this model.
\end{abstract}

{\bf Keywords: Entanglement classification; Data structures; Quantum protocols}

\section{Introduction}\label{sec: introduction}
Recently the field of quantum information and computation (herein QIC), due to technological improvements, has gained increasing interest. In particular, in computer sciences, researchers are investigating whether abstract models and logical reasoning may be applied to design and to formally verify quantum protocols and quantum dynamical systems. 

Quantum systems show behaviors different from the classical ones, among those are superposition, interference, and \emph{entanglement} \cite{HHHH}. Some of the aforementioned behaviors are manifested even in one-particle effects. Some other effects, such as entanglement, can be observed only in composite quantum systems, i.e., systems composed of two or more subsystems. These subsystems can exhibit both classical and non--classical correlations. \emph{Entanglement} is a kind of non--classical correlation, displayed by certain classes of quantum states called \emph{entangled states}, and it is a vital concept in QIC. It constitutes a fundamental resource for many quantum protocols --ranging,  e.g., from the simplest case of quantum teleportation\cite{teleport1} to more complex scenarios such as quantum key distribution (herein QKD) and quantum cryptography\cite{qkd}-- which strongly rely on entanglement to be properly performed. In order to perform formal reasoning in quantum protocols, entanglement  \emph{as a resource} needs to be classified, since each \emph{entanglement class} is associated with a different set of tasks in quantum information processing. However, this classification is not trivial, as the notion of entanglement differs according to the number of entangled particles of the quantum system under consideration.

In \emph{bipartite entanglement}, one deals with quantum systems which are composed by two subsystems. Thus, a bipartite quantum system is associated to the Hilbert space $\mathcal{H}=\mathcal{H}^{(1)}\otimes\mathcal{H}^{(2)}$\footnote{$\mathcal{H}^{(i)}$ are the Hilbert spaces associated to each single particle.} and, given two states $|\psi_1\rangle \in \mathcal{H}^{(1)}$ and $|\psi_2\rangle \in \mathcal{H}^{(2)}$,  one can always build a state $|\Psi_{sep}\rangle= |\psi_1\rangle \otimes |\psi_2\rangle \text{,}\ \ |\Psi_{sep}\rangle \in \mathcal{H}$. Quantum states that can be written in this form are called \emph{separable} (or product states). It is important to recall that not every state in $\mathcal{H}$ is separable; there are, indeed, bipartite quantum states that cannot be decomposed as above and hence are called entangled.

Quantum entanglement in bipartite systems of pure states --i.e., states corresponding to vectors in a Hilbert space--  is (almost) completely understood, since it can be characterized by using the \emph{Schmidt decomposition} \cite{Schmidt}. This decomposition allows one to write down, by means of local unitary transformation only, any pure state $|\Psi \rangle$ of a bipartite system in the canonical form displayed in Eq.(\ref{Schmidt}):
\begin{equation}
|\Psi\rangle= \sqrt{\lambda_i}\sum_i |\psi_i\rangle\otimes|\varphi_i\rangle\, ,
 \label{Schmidt}
\end{equation}
where $\lambda_i$ are called \emph{Schmidt coefficients} and the local bases $\{|\psi_i\rangle\}$ and $\{|\varphi_i\rangle\}$ are guaranteed to always exist. The sum is limited by the dimension of the smaller Hilbert space, i.e., $\mathcal{H}^{(1)} $ or $\mathcal{H}^{(2)}$. The non--local properties of the state are encoded in the positive Schmidt coefficients, which tell us whether the state is separable or not. In fact, if at least two Schmidt coefficients are different from zero, the state is not expressible as a product, meaning that it is entangled. 

Quantum entanglement in multipartite systems of pure states  is not as easy to classify as in the bipartite case. Indeed, when addressing \emph{multipartite entanglement}, which refers to correlations between more than two subsystems, it is not enough to know whether the subsystems are entangled or not, but also \emph{how} entangled they are. There are different ways in which a pure state $|\Psi\rangle \in \mathcal{H}^{(1)} \otimes \dots \otimes \mathcal{H}^{(N)}$ in an $N$-partite system can be entangled. In particular, in tripartite systems of qubits --i.e., systems represented by the Hilbert space $\mathbb{C}^2 \otimes \mathbb{C}^2 \otimes \mathbb{C}^2$--  there are separable states, biseparable states and two types of fully inseparable states, i.e., Greenberger-Horn-Zeilinger (herein GHZ) states \cite{GHZ} which have the form $|GHZ\rangle = \frac{1}{\sqrt{2}}(|000\rangle+|111\rangle)$ and  $W$-states, which are entangled states of three qubits of the form $|W\rangle=\frac{1}{\sqrt{3}}(|001\rangle+|010\rangle+|100\rangle)$. $GHZ$ and $W$ states are locally inequivalent, i.e., they cannot be transformed in each other by means of stochastic local operations assisted by classical communication (herein SLOCC) \cite{DVC}.

In the case of multipartite entanglement, as in the bipartite ones, there exists a \emph{generalized Schmidt decomposition} (herein GSD) \cite{Acin, Sudbery} which will be used by the classification approach considered in this work. GSD allows us to put any state in a canonical form, which is then used as a starting point for other algorithmic steps, involving entanglement measures. We use \emph{concurrence} \cite{Wootters} as the measure to quantify bipartite entanglement in multipartite systems, as defined in Eq.(\ref{concurrence}): 
\begin{equation}\label{concurrence}
\mathcal{C}(\rho)=\max\left\{0,\sqrt{\mu_{1}}-\sqrt{\mu_{2}}-\sqrt{\mu_{3}}-\sqrt{\mu_{4}}\right\}\, ,
\end{equation}
where the $\mu_{i}$ are the non-negative eigenvalues, in decreasing order, of the non-Hermitian matrix $\rho{\tilde \rho}$. Here, $\tilde{\rho}$ is the matrix given by $\tilde{\rho}=\left(\sigma_{y}\otimes\sigma_{y}\right)\rho^{*}\left(\sigma_{y}\otimes\sigma_{y}\right)$ where $\rho^{*}$ is the complex conjugate of $\rho$ \footnote{i.e., the density operator associated to the quantum state under consideration, according to its usual definition} when it is expressed in a standard basis such as $\{|00\rangle, |01\rangle, |10\rangle, |11\rangle\}$ and $\sigma_y$ represents the Pauli $Y$ operator. In this paper, since we focus on the case of three-qubit systems, we use \emph{tangle} \cite{CKW} to assess tripartite entanglement. In fact, in tripartite systems, tangle allows the quantification of entanglement between states that are not pairwise entangled. However, tangle cannot properly quantify the tripartite entanglement of $W$ state since $\tau(|W\rangle)=0$ but it still remains a well-known quantity that distinguishes GHZ-type (which have non-vanishing tangle) and $W$-type states. Given $\mathcal{H}=\mathbb{C}^2_A\otimes\mathbb{C}^2_B\otimes \mathbb{C}^2_C$, tangle is defined as in Eq.(\ref{tangle}):
\begin{equation}\label{tangle}
\tau=\mathcal{C}_{A(BC)}^{2}-\mathcal{C}_{AB}^{2}-\mathcal{C}_{AC}^{2}\, ,
\end{equation}
where the bipartite concurrence $\mathcal{C}_{A(BC)}$ is defined as $\mathcal{C}_{A(BC)}=\sqrt{2\left(1-{\rm Tr}(\rho_{A}^{2})\right)}$, and ${\rm Tr}$ represents the \emph{trace} function in its usual definition.

Understanding \emph{how} multipartite systems are entangled is a formal problem that, if solved efficiently, has positive impact on many applications. In order to do that, we aim at building a finite classification of entangled systems of three qubits, by now restricted to the tripartite case, using methods that can be further investigated  from a logical and computational point of view  as well (e.g., graphs and hypergraphs). 
Since entanglement is an important resource in QIC, its classification can be useful to verify properties on quantum protocols, in particular in the context of quantum cryptography. Formal techniques, such as \emph{model--checking}, can be used to test the reliability of a quantum protocol in a realistic scenario, thus an approach allowing us to have a finite classification of quantum entanglement in the multipartite case can be a further step in the direction of building, and then testing, realistic  protocols.

In this work, we extend the idea presented in Ref. \cite{GA} which was in turn based on the concept of entangled graph in Ref. \cite{PB}. The paper is structured as follows: Section \ref{sec: tfpartite} explores, in details, a method to classify three-qubit entanglement by means of a new concept called \emph{entangled hypergraph} (herein EH), in order to algorithmically build the classification. We aim to extending this approach to the multipartite case, but this problem is currently still under investigation. Section \ref{sec: EEH} uses the proposed classification to build a new data structure, called \emph{evolving entangled hypergraph} (herein EEH), suitable to represent evolving quantum systems in which entanglement is an emergent behavior. Section \ref{sec: examples} shows some examples of the proposed method, applied to quantum teleportation and QKD protocol which use entanglement. Section \ref{sec: conclusion} ends the paper by exploring future ideas and implementation proposals.

\section{Classification of Three-qubit Entanglement}\label{sec: tfpartite}
In this section, we provide a classiﬁcation for three-qubit pure states inspired by the approach presented in Ref. \cite{GA}. The classification proposed in this work uses the notion of GSD of three-qubit pure states, together with the concept of {\it entangled hypergrap} which can be considered as a generalization of entangled graph. In fact, we can define the latter as follows:

\textbf{Definition 1 (Entangled Graph).} An entangled graph is an ordered pair $G=(V,E)$ in which each vertex $v_i \in V$ represents a qubit in multi-qubit system, and an edge $e(v_i,v_j) \in E$ between two different vertices denotes bipartite entanglement between the corresponding qubits\cite{PB}.

\begin{figure}[h!]
\centerline{\includegraphics[height=7cm]{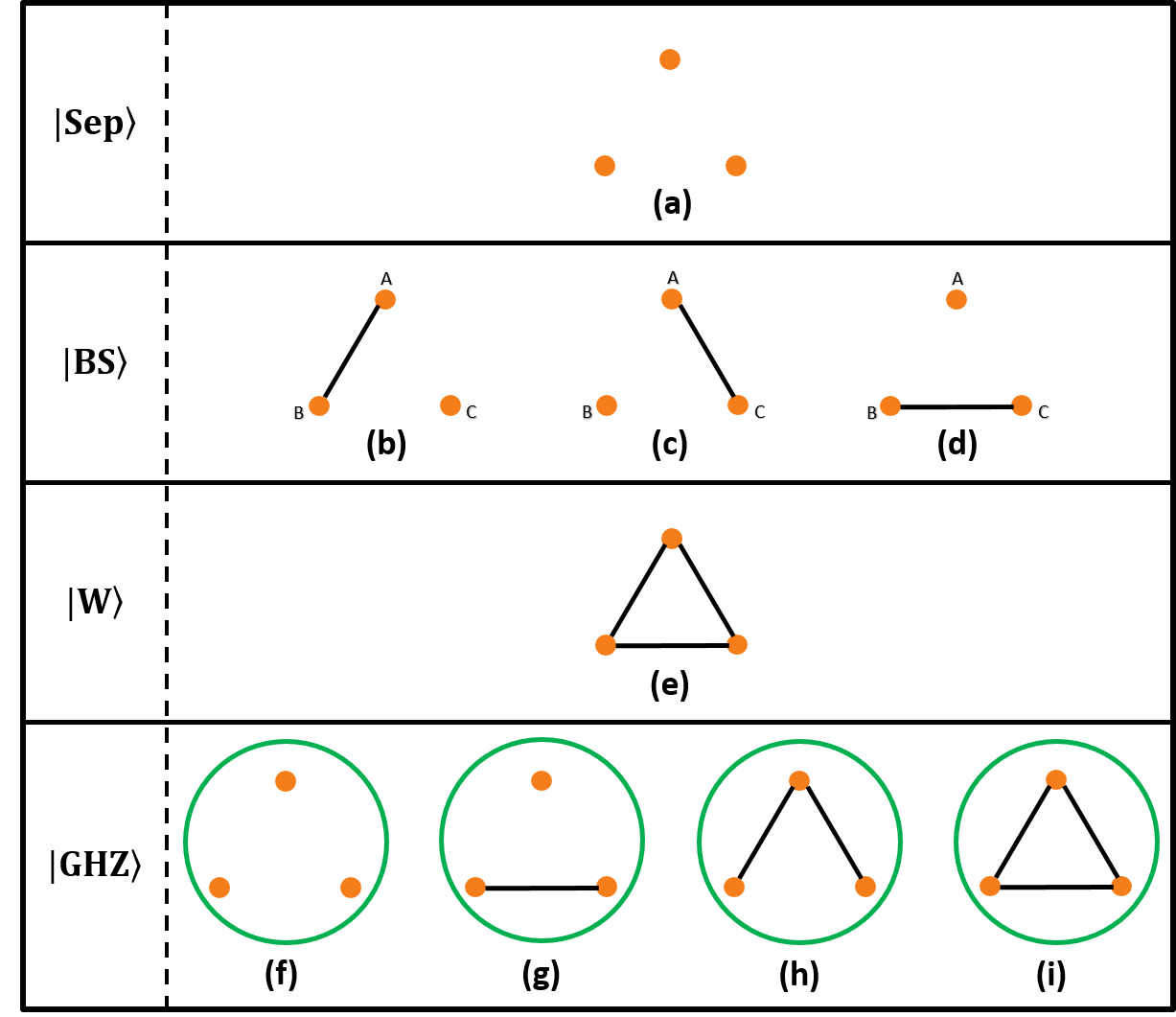}}
\caption{Classification of three-qubit entanglement in terms of entangled hypergraphs}
\label{figure:1}
\end{figure}

However, by using the concept of entangled graph we can only assess bipartite entanglement in multi-qubit systems \cite{PB2}. To overcome this problem, in Ref. \cite{GA}, the authors used a circle enclosing the graph indicating global entanglement in order to avoid ambiguity between corresponding GHZ and separable states in the same entangled graph. It is important to note that, when we consider systems composed by four (or more) qubits, it is not enough to investigate only bipartite and global entanglement, thus it would be more beneficial to use another data structure. For this reason we exploit the concept of entangled hypergraph. According to this very consideration, we decided to associate an entangled hypergraph to each possible entangled state (see Fig. \ref{figure:1}).

\textbf{Definition 2 (Entangled Hypergraph).}
An entangled hypergraph is an ordered pair $G=(V,H_G)$ in which each vertex $v_i\in V$ represents a qubit of a multi-qubit system and each hyperedge $h_{k}(v_i,v_j,\dots,v_m) \in H_G$ (also know as $k$-edge) links $\text{k}\geq 2$ vertices, indicating $k$-partite entanglement between the corresponding qubits.

In Ref. \cite{Acin}, a GSD is presented such that, for every three-qubit pure state, there exists a local base allowing to rewrite the state in a unique canonical form, by using a set of five orthogonal product states. This can be expressed as in Eq. (\ref{GSD3}):
\begin{eqnarray}\label{GSD3}\nonumber
|\Psi\rangle_{3}&=&\lambda_{0}|000\rangle + \lambda_{1} {\rm e}^{i\theta}|100\rangle + \lambda_{2}|101\rangle + \lambda_{3}|110\rangle + \lambda_{4}|111\rangle\, , \\
& & \lambda_i\ge 0\, , \quad 0\le \theta\le \pi\, , \quad \sum_{i}\lambda_i^2=1\, .
\end{eqnarray}
As a matter of fact, it can be proven that any pure three-qubit state is locally unitary equivalent to this form. 

The first step of our approach is thus to compute the GSD of the state under consideration.
Now, let us consider all the possible bipartite factorizations with nonzero concurrence of this state, i.e., $\mathcal{C}_{ij}\neq 0 \ \ \forall i,j$ with $ij$ referring to the $i$-th and $j$-th qubits. This allows us to find coefficients of the GSD which make nonzero concurrences and which guarantee a weighted edge (i.e., $2$-edge) between two vertices. Then, in addition, we also need to consider tripartite entanglement, which corresponds to global entanglement in the case of tripartite systems. Hence, we use the tangle and find coefficients of the GSD which make nonzero tangle, i.e., $\tau \neq 0$, providing a weighted 3-edge. These steps are summarized in Table \ref{table:1}. 

\begin{figure}[h]
\centerline{\includegraphics[height=3.2cm]{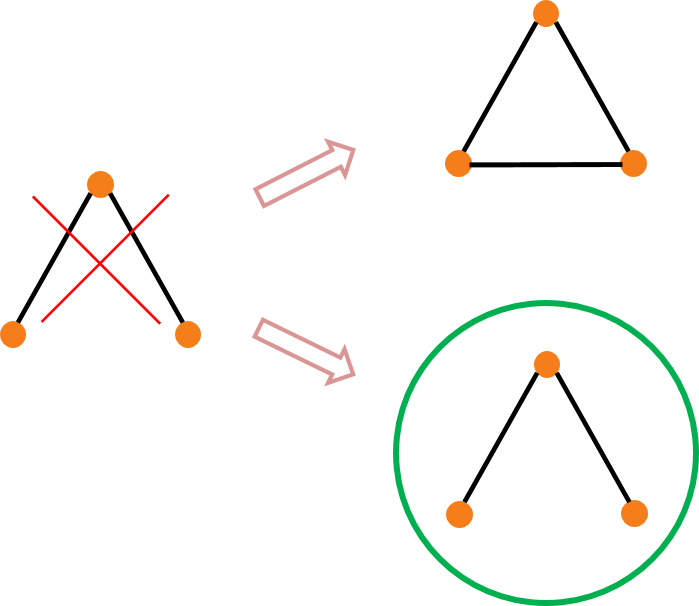}}
\caption{Forbidden entangled hypergraph of three-qubit entanglement}
\label{figure:2}
\end{figure}

Permutations of entangled hypergraphs can be considered by labeling the vertices. For instance, in Figure \ref{figure:2} we have labeled the vertices of entangled hypergraphs in the biseparable case. In this way, we can have a relation between this classification and the SLOCC classification, since we know there are three different SLOCC classes of biseparable states, where all of them belong to the same family. 
Since the entangled hypergraphs corresponding to separable and W states are symmetric, i.e., permutation-wise invariant, and the ones corresponding to GHZ-type states contain an hyperedge, we have not labeled their vertices.

It is important to notice that among all the possible hypergraphs, there are those that are not corresponding to any entangled pure state  (i.e., starting from the hypergraphs, it is not possible to retrieve the associated state). We call this kind of structures \emph{forbidden entangled hypergraphs}. In the three-qubit case, the forbidden entangled hypergraph is the one with two edges but no hyperedge, i.e., with no global entanglement. Indeed, for every possible choice of the coefficients to have two edges, we always end up having either the third edge or the hyperedge (see Fig. \ref{figure:2}).

For 4 qubits or more, we have infinite number of SLOCC classes so it is desirable to group the infinite number of classes to a finite number of families. According to the forbidden entangled hypergraph in three-qubit case, there are two genuine entangled families. We guess the forbidden entangled hypergraphs are hypertrees but investigating this problem is out of the scope of this work. In this manner, we can extend this entanglement classification at least for a finite number of qubits.

\begin{table}
\centering
\caption{Three-qubit GSD's coefficients that make nonzero concurrences and tangle}
\vspace{8pt}
\begin{tabular}{c c c c}
\hline
& & & \\ [-2.5ex]
$\mathcal{C}_{12}$ & $\mathcal{C}_{13}$ & $\mathcal{C}_{23}$ & $\tau$ \\ [0.5ex]
\hline
& & & \\ [-2.5ex]
($\lambda_{0}$ , $\lambda_{3}$) & ($\lambda_{0}$ , $\lambda_{2}$) &
($\lambda_{1}$ , $\lambda_{4}$) & ($\lambda_{0}$ , $\lambda_{4}$) \\ [0.5ex]
& & ($\lambda_{2}$ , $\lambda_{3}$)\\ [0.5ex]
\hline
\end{tabular}
\label{table:1}
\end{table}

\section{Evolving Entangled Hypergraphs}\label{sec: EEH}

The core part of this work is devoted to use the classification introduced in Section \ref{sec: tfpartite} as a starting point to model quantum protocols which use entanglement as a resource. We propose the concept of \emph{evolving entangled hypergraph} (herein EEH), a data structure inspired by both concepts of entangled hypergraphs and hypergraph states\cite{QHS}, taking into account not only correlations (i.e., entanglement) but also interactions and the time evolution of the system under consideration. Interactions in this case are considered in the context of a dynamical process, i.e., the evolution of the quantum system in successive, discrete time steps. Since we like to keep track of patterns in the underlying hypergraphs, we decided to create a structure that merged the notion of entangled hypergraphs together with the one of  \emph{multilayer graphs}.  This approach turns out to be useful, in particular, when we try to model quantum protocols in which entanglement is an important resource to be preserved, e.g., QKD and quantum teleportation.

Multilayer graphs belong to the family of evolving graphs, also known as temporal graphs \cite{6}. 
Evolving  graphs highlight the change  in  time of a graph.  According to Ref. \cite{6.1}, if in an evolving graph $G=(V,E)$ the time is understood in a discrete manner and only the relationships (i.e., edges) between entities (i.e., nodes) may change leaving the graph topology unchanged,  then $G$ is a  sequence $G_1, \dots , G_n$ of static graphs over the same set of nodes. Evolving graphs \cite{6.2} have also been proposed as a  theoretic model in order to capture the changes in time of a dynamic network topology. According to Ref. \cite{6.3}, these graphs are suitable to analyze the quality of a communication protocol, since its \emph{history} is made explicitly of a sequence of graph topologies.  
A \emph{multilayer graph} \cite{6.4} is an evolving graph made of layers which are distinct copies of the main \emph{spatial} graph, i.e., a graph in which nodes or edges are spatially located according to a certain metric. Each layer is then connected to its neighbor according to some measure of time (e.g., causality, probability, etc.) which encodes the evolution of the network. 

The multilayer graphs approach is thus suitable to describe the behavior of a dynamic system taking into account both spatial and temporal dimensions. For this reason, we decided to use EEHs as a method to represent quantum dynamics in which entanglement is an emergent behavior. 

\textbf{Definition 3 (Evolving Entangled Hypergraph).}
Let $G=(V, H_G)$ be an entangled hypergraph. 
An evolving entangled hypergraph $EEH=(L, H)$ is an evolving multilayer graph in which:
\begin{enumerate}
\item $L=\{L_0,\dots, L_{t-1}\}$, with the variable $t$ representing the time-steps, is a set of \emph{layers}. Each layer $L_i \in L$  represents an instance of $G$ at time $t_i$;
\item $H=\{H_1,\dots, H_{t-1}\}$  is the set of hyperedges from a layer to the following one. Each hyperedge $h_i \in H$ is labeled with a CPTP (i.e., completely positive and trace preserving) linear map acting on the states of G. 
\end{enumerate}
In other words, $H$ contains edges  from a layer to the following layer, which are called \emph{inter-layer} hyperedges, while edges within a single layer $L_i$ are called \emph{intra-layer} hyperedges.

\begin{figure}[h]
\centerline{\includegraphics[width=\textwidth]{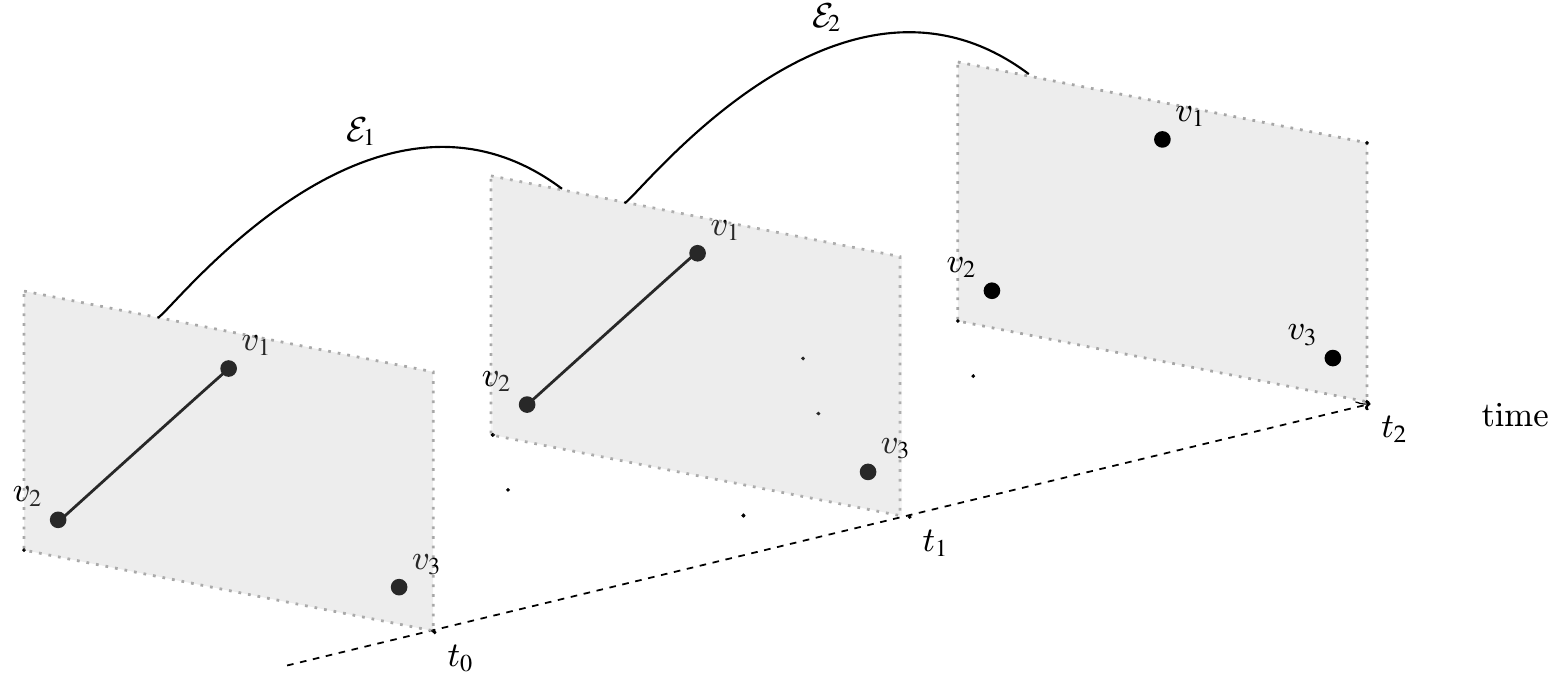}}
\vspace*{8pt}
\caption{Example of an EEH in which decoherence occurs.}\label{fig: multilayer}
\end{figure}

\begin{figure}[h]
\centerline{\includegraphics[width=\textwidth]{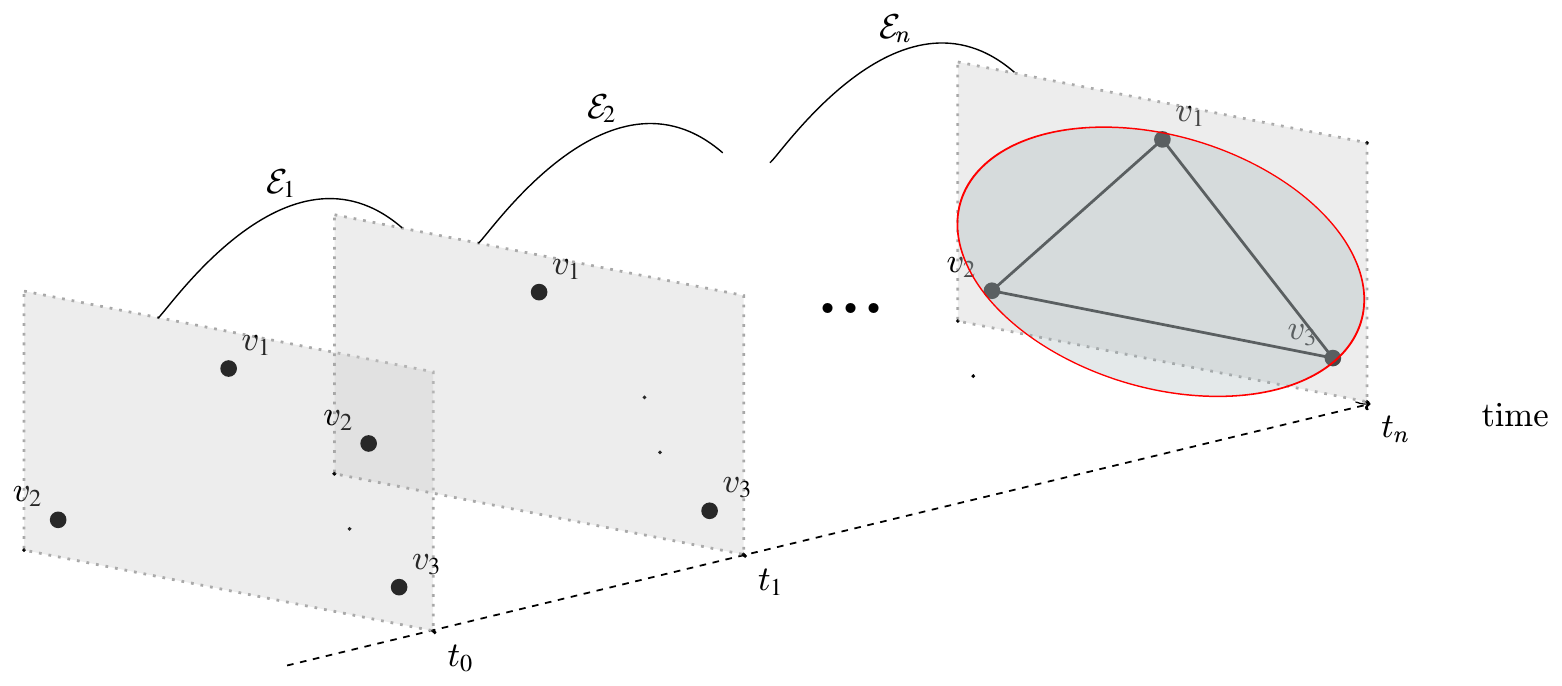}}
\vspace*{8pt}
\caption{Example of an EEH in which entanglement creation occurs.}\label{fig: multilayer2}
\end{figure}

\noindent Each $L_i$  corresponds to one of the allowed EHs, belonging to the classification presented in Section \ref{sec: tfpartite} (i.e., one of the allowed equivalence classes). 
Figures (\ref{fig: multilayer}, \ref{fig: multilayer2}) show two examples of  hypothetical EEHs in a tripartite system. It is important to note that the representation of states is given in the density matrix formalism, since it allows a broader set of unitary and non-unitary evolution operators. 
In Fig.\ref{fig: multilayer}, an initial state $\rho$ (i.e., a given state of three qubits) at time $t_0$ is biseparable and,  after the evolution through two given CPTP maps  --$\mathcal{E}_1$ and $\mathcal{E}_2$-- ends in a completely separable state at time $t_2$. The information extracted from this (hypothetical) model reveals that, during the execution of the quantum protocol under consideration, in the quantum channel $\mathcal{E}_2$  something occurred that causing decoherence. 

In the example in Fig. \ref{fig: multilayer2}, we show an EEH  in which the state $\rho$ is in a completely separable configuration at time $t_0$. Then,  after the evolution through the maps $\mathcal{E}_1, \dots, \mathcal{E}_n$, at time $t_n$ it becomes totally entangled, allowing us to witness a process of entanglement creation along the channel.

\noindent \emph{Intra-layer} edges  $e_i\in E$ are not labeled; a weight can eventually be added, referring to the amount of entanglement between two qubits, quantified by concurrence, or by  von--Neumann entropy. Nodes $v_i \in V$ in the EH (i.e., within a layer) are labeled and should be interpreted as spatial locations of qubits (i.e., qubit positions) instead of qubit ``\emph{names}". We suggest that this subtle difference, even if not used in this early stage of the work, might become more important when dealing with identical particles, thus the notion of \emph{indistinguishability} will be enforced by using locations of qubits instead of labeling them.

The EEH model proposed in this work is an \emph{abstract} structure taking as input a class $L_i$ from a finite set $L$ of equivalence classes, which evolves according some CPTP map $\mathcal{E}$ and producing, as output, another class from the same set; i.e., $\{L_i\}\xrightarrow{\mathcal{E}}\{L_j\}$. 
The main purpose of EEHs is to create a structure that models properties emerging from quantum dynamics like, in this case, entanglement. Such properties should be suitable to verify using formal methods such as model--checking and logic, thus it is important to keep EEHs as abstract as possible. This is useful since by abstracting the domain, we ``forget" some details, making the model computer representable. 

We shall not aim at computing the numerical representation of the operators which are labeling the inter-layer hyperedges. In this way, the formal verification tool  should rely not on an explicit  --and computationally expensive-- representation of the CPTP maps, but instead it focuses on the abstract representation of their action on the graph. Using this approach, which can intuitively be regarded as an \emph{abstract interpretation} one, a formal verification tool gains information about the semantics and the properties of the quantum protocol under consideration, without performing calculations.

In the EEHs, each layer representation (i.e., sub-hypergraph G) encodes an action induced by the CPTP map entering it. In this way, a quantum channel (i.e., CPTP map) has a representation at the hypergraph level and its behavior in time removes (or adds) \emph{intra-layer} edges. The rules determining how many edges are allowed to be removed (or added) relates to the classification presented in the first part of this work.

\section{Example: Modeling Quantum Protocols}\label{sec: examples}
In this section we show an application for EEHs by using them to model two quantum protocols, i.e., teleportation and QKD with $W$ states.
We suggest that EEHs are suitable to model quantum dynamics in which entanglement is a property to be preserved or detected, because their structure allows to track the structural and morphological evolution encoding both spatial and temporal behaviors of an evolving quantum system.
EEHs are abstract structures, thus they allow to remove unwanted, and computationally expensive, information by focusing on the structure of the hypergraph. Moreover, they can be constructed by using an algorithmic approach.
 
In the following we will provide two instances of  ``real world" quantum teleportation protocols in which tripartite entanglement is used as a resource and for which EEHs provide a good model. We represent the protocols by using both a Quipper-like \cite{quipper} pseudocode and a graphical representation of the associated EEH.

It is also important to stress that, since this is just a preliminary work, we do not consider measurements in detail, which will be further investigated in the extension of this paper. Instead, in this work we will focus our attention just on the channel.

\subsection{Quantum Teleportation Protocol}
Quantum teleportation is a protocol allowing to transmit quantum information (e.g. a quantum state) from one location to another, using both classical communication and quantum entanglement between the sender and the receiver. In this context we abstract from the underlying physical and mathematical details, since they are out of the scope of this preliminary work; further references can be found in Ref. \cite{teleport1}.

\begin{figure}[h]
   \begin{minipage}{0.48\textwidth}
     \centering
\begin{lstlisting}[language=Haskell, breaklines=true, firstline=1, basicstyle=\scriptsize\ttfamily\linespread{0.5}, lastline=20]
teleport :: (Qubit, Qubit, Qubit) -> Qubit
teleport (q1, q2, q3) :: do
reset_at (q2,q3)
EPR_at (q2,q3)
CNOT_at q2 controlled q1
hadamard_at q1
c1 <- measure q1
c2 <- measure q2
if c1==1 && c2==1
        then do
          gate_Y_at q3
        else if c1==1 && c2==0
          then do
            gate_Z_at q3
          else if c1==0 && c2==1
            then do
              gate_X_at q3
            else do
              identity_at q3
return q3
\end{lstlisting}
   \end{minipage}\hfill
   \begin {minipage}{0.48\textwidth}
     \centering
\includegraphics[scale=0.80]{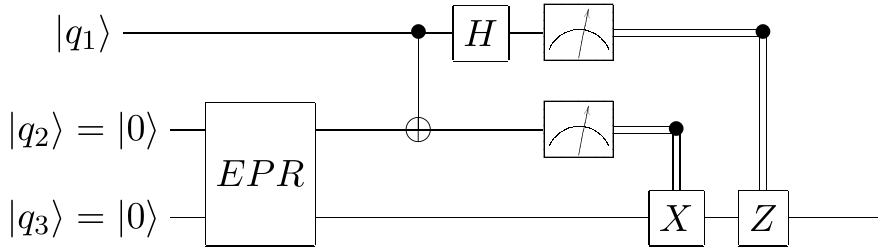}
   \end{minipage}
\protect\caption{Quantum Teleportation Pseudocode and Circuit}
\label{fig:example_1}
\end{figure}

In Fig. \ref{fig:example_1} we have provided a pseudocode implementation for the teleportation protocol together with its circuit representation, where the \texttt{EPR} gate creates an entangled (e.g., Bell) state of the form $\frac{1}{\sqrt{2}}|00\rangle+|11\rangle$ and \texttt{CNOT} represent the standard \emph{controlled-not} gate. 

\begin{figure}[h]
\centerline{\includegraphics[width=\textwidth]{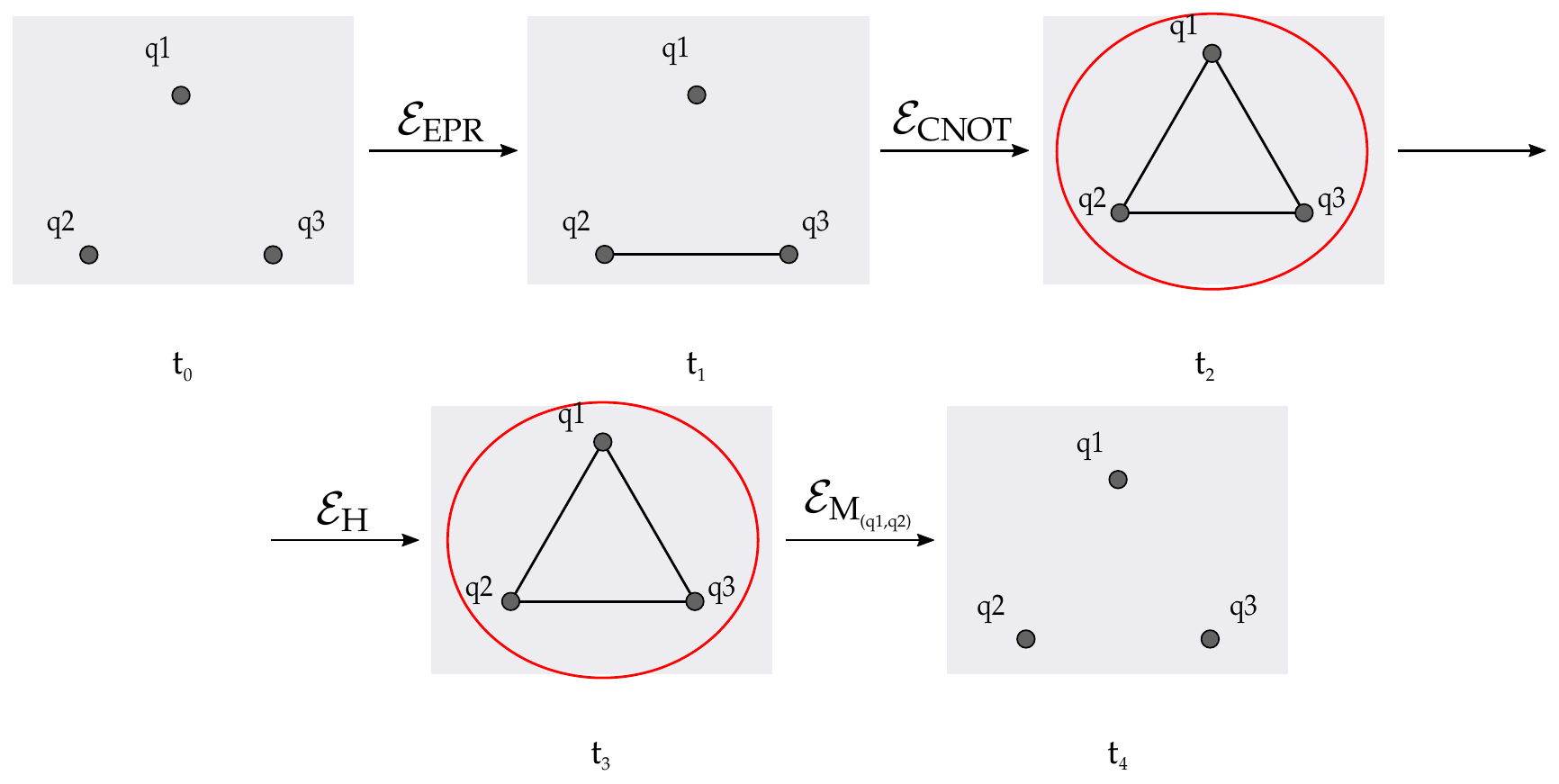}}
\vspace*{8pt}
\caption{EEH for an ideal quantum teleportation protocol.}\label{fig: nonoise}
\end{figure}

In the case in which no noise along the channel occurs, this protocol --without considering  measurement and post measure corrections-- can be summarized as in Eq. \ref{eq:ent} (note that normalization factors are omitted for simplicity):

\begin{align}&\ \ \ \ \ \ \ \ \ |q_1\ 0\ 0\rangle \ \ \ \ \ \ \ \ \ \ \ \ \ \ \ \ \footnotesize\text{separable}\nonumber \\ &\xrightarrow{EPR} |q_1\rangle(|00\rangle+|11\rangle) \ \ \ \ \ \footnotesize\text{biseparable}\nonumber \\ &\xrightarrow{CNOT \ } |0\rangle(|00\rangle+|11\rangle)+ |1\rangle(|10\rangle+|01\rangle) \ \ \ \ \ \footnotesize\text{fully entangled}\nonumber \\
&\xrightarrow{\ H\ \ } |+\rangle(|00\rangle+|11\rangle)+ |-\rangle(|10\rangle+|01\rangle) \ \ \ \ \footnotesize\text{fully entangled}\label{eq:ent}
\end{align}
\noindent where $|\pm\rangle$ represent the states $|0\rangle \pm |1\rangle$. It is important to note that the application of a controlled-not gate allows to ``extend" the entanglement to the first subsystem creating a fully entangled state which can be then reduced to a $GHZ$-like state with both bipartite and tripartite entanglement, according to the underlying classification. \footnote{The calculations have been performed by using the QETLAB\cite{qetlab} toolkit for MATLAB, computing both von-Neumann entropy --function \texttt{Entropy}-- and concurrence --function \texttt{Concurrence} of the subsystems.}
This process results in the EEH in Fig. \ref{fig: nonoise}, in which the quantum gates are represented by $\mathcal{E}_{\text{gate-name}}$, i.e.,  the quantum operations associated the circuit gate. This choice allows to deal with more realistic scenarios, in which noise and decoherence (i.e., processes that are not unitary) may occur.

\begin{figure}[h]
   \begin{minipage}{0.48\textwidth}
     \centering
\begin{lstlisting}[language=Haskell, breaklines=true, firstline=21, basicstyle=\scriptsize\ttfamily\linespread{0.5}, lastline=40]

teleport :: (Qubit, Qubit, Qubit) -> Qubit
teleport (q1, q2, q3) :: do
reset_at (q2, q3)
EPR_at (q2,q3)
PhaseFlip_at (q1,q2,q3)
CNOT_at q2 controlled q1
hadamard_at q1
c1 <- measure q1
c2 <- measure q2
if c1==1 && c2==1
        then do
          gate_Y_at q3
        else if c1==1 && c2==0
          then do
            gate_Z_at q3
          else if c1==0 && c2==1
            then do
              gate_X_at q3
            else do
\end{lstlisting}
   \end{minipage}\hfill
   \begin {minipage}{0.48\textwidth}
     \centering
\includegraphics[width=\textwidth]{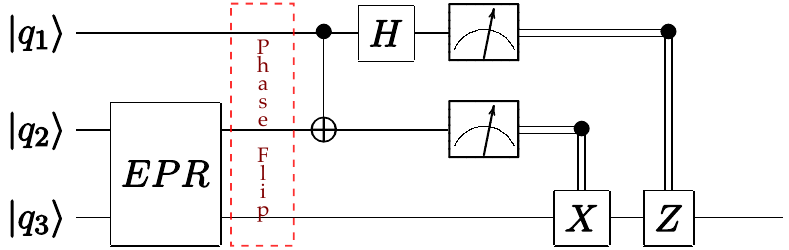}
   \end{minipage}
\protect\caption{Teleportation Pseudocode and Circuit with Noise}
\label{fig:example_2}
\end{figure}

Let us now consider the same teleportation protocol in which we add a \emph{phase flip} channel instead of an ideal, not noisy one. This simulates a situation in which a loss of coherence may occur and can be realized by the  pseudocode in Fig. \ref{fig:example_2}, in which \texttt{PhaseFlip\_at} represents a phase flip channel.

\begin{figure}[h]
\centerline{\includegraphics[width=\textwidth]{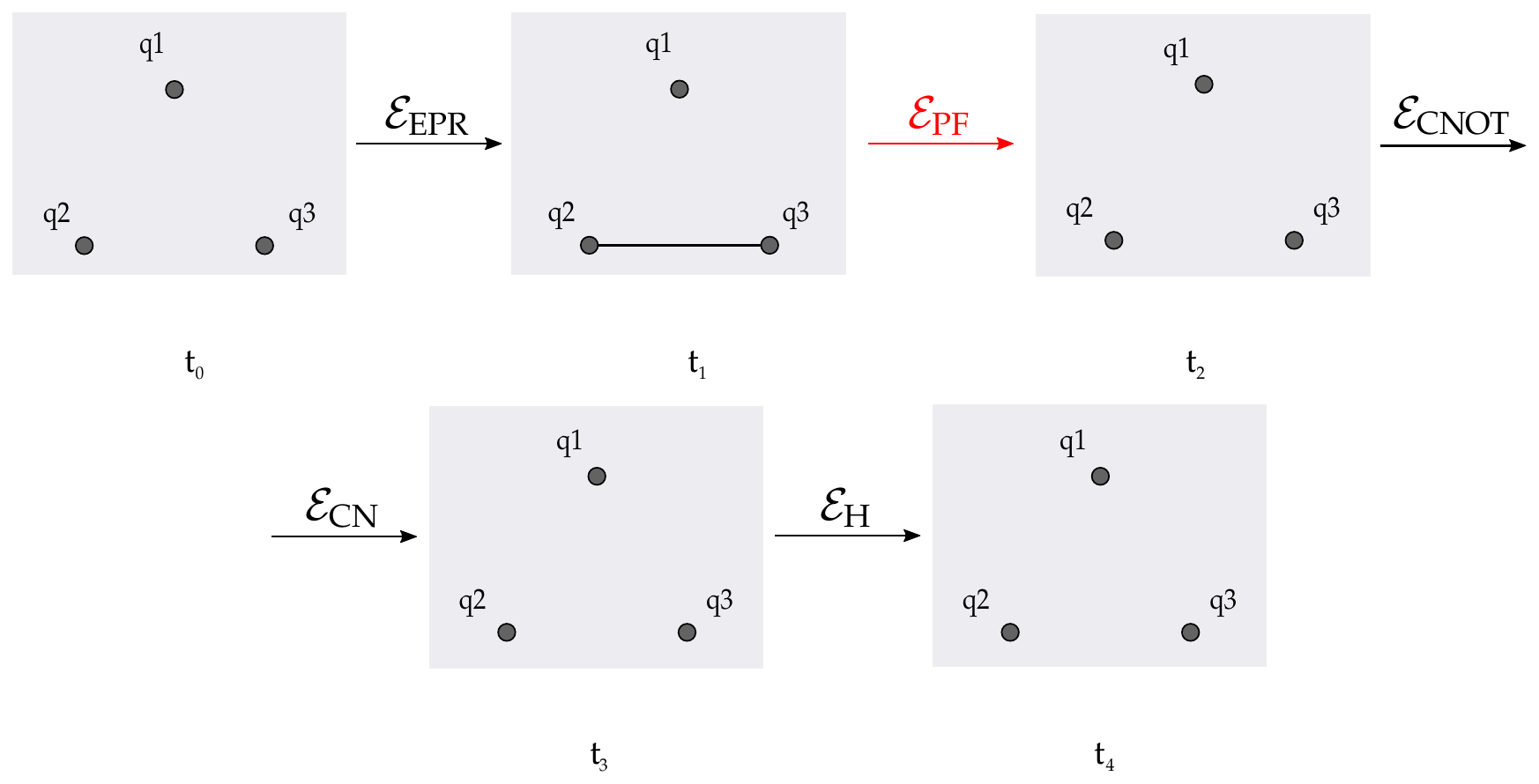}}
\vspace*{8pt}
\caption{EEHs for a teleportation protocol with noise.}\label{fig: noise}
\end{figure}

Given a generic state $\rho$, the phase flip channel is represented by the quantum operation $\rho '=\sum_{i=1}^2 E_i \rho E_i^{\dag}$ , with $E_1=\sqrt{p}\sigma_z$, $E_2=\sqrt{1-p}\mathbb{I}$ and $0\leq p\leq 1$. \footnote{$\sigma_z $ is the Pauli Z operator, applied to the whole system and $\mathbb{I}$ the identity matrix.} After the application of the channel, the sum of the coherences (i.e.,  the off-diagonal) of the density matrix $\rho '$ is lower than in the initial state; by iterating the process $k$ times, the state loses all the information about coherences (i.e., decoherence), becoming less entangled with the rest of the system at each iteration. In this case we suppose that the probability $p$ is high enough to guarantee a complete decoherence with just one application of the channel. Thus, the EEH associated is the one in Fig.\ref{fig: noise}. The hypergraph is different from the previous one in two layers, namely $t_2$ and $t_3$, and since the entanglement is not modelled in the same way as before, it is not possible to guarantee the effectiveness of the protocol, which could fail to teleport the correct state. 

Different teleportation methods, i.e., through quantum channels containing more than two entangled qubits, have been investigated. In  Refs. \cite{teleport2, teleport3} tripartite GHZ and $W$ states can also be used as quantum channels for teleportation, which can be an interesting case study, since they deal with more than three qubits.

\subsection{QKD Using W States}
In this example we consider a quantum key distribution protocol using $W$ states, as presented in Refs. \cite{w1,w2}.
$W$ states, due to their pairwise entanglement, have been considered suitable and robust configurations for QKD protocols. Indeed, after tracing out one subsystem, there is still the possibility of bipartite entanglement, while the GHZ state, once that a subsystem have been traced away, becomes completely separable.
The QKD protocol via $W$ states between three parties can be summarized as follows:
\begin{enumerate}
\item The three parties share respectively one qubit each, which belongs from a previously entangled  tripartite $W$ state;
\item They randomly choose a basis to locally measure their qubit (e.g., $x$ or $z$-basis);
\item Each part announces a bit of information on the basis of the local measurement (not the outcome);
\item For security reasons there might be a requests to announce the outcomes at random, to discover if previously eavesdropping has occurred;
\item If  the three measurement basis are not  $(z,x,x)$, $(x,z,x)$ or $(x,x,z)$, then the protocol is restarted;
\item If the outcome of the part who measured in the $z$-basis is $|0\rangle$ then the protocol ends and the other two parties know for sure that they have the same outcome, otherwise the protocol is restarted.
\end{enumerate}

\begin{figure}[h]
     \centering
\begin{lstlisting}[language=Haskell, breaklines=true, firstline=44, basicstyle=\scriptsize\ttfamily\linespread{0.5}, lastline=56]

qkdW :: (Qubit, Qubit, Qubit) -> (Qubit, Qubit, Qubit)
qkdW (q1, q2, q3) = do
    W_at (q1, q2, q3)
    A <- measure_x-z q1
    B <- measure_x-z q2 
    C <- measure_x-z q3
    if (axis(A,B,C)==(z,x,x) || axis(A,B,C)==(x,z,x) || 
       axis(A,B,C)==(x,x,z)) && C == |z+>
              then do
              return (m1,m2,m3)
              else qkdW(q1, q2, q3)
\end{lstlisting}

\begin{center}
\includegraphics[width=\textwidth]{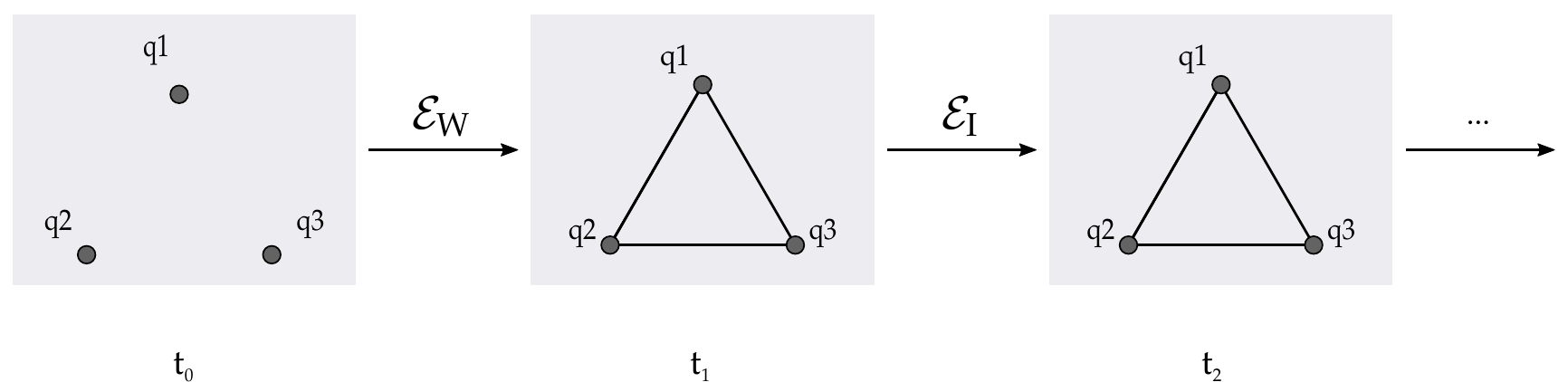}
\protect\caption{QKD with $W$ states pseudocode and EEH.}
\label{fig:wnonoise}
\end{center}
\end{figure}

This protocol heavily relies on the preservation of the entanglement of the $W$-states. If the entanglement is preserved, then the protocol ends in a success, otherwise, if noise along the channel or eavesdropping occur before the measurement, the protocols ends in a failure and should be restarted. For this reason we will focus only on the part of the protocol in which the entangled state is generated and transmitted along the channel, i.e., the first point of the above list. In Fig. \ref{fig:wnonoise} we provide the pseudocode implementation  for the QKD protocol via $W$ states, together with the portion of EEH relative to the pre-measurement part, where the identity operator $\mathcal{E}_I$ refers to a channel without noise.

\begin {figure}[h]
\begin{center}
\includegraphics[width=\textwidth]{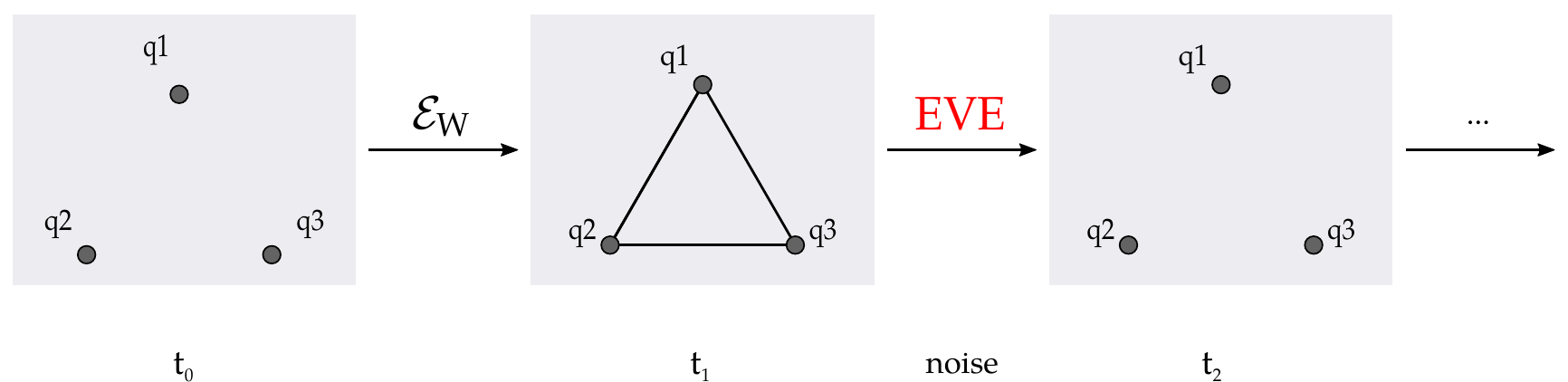}
\protect\caption{QKD with $W$ states EEH noisy channel.}
\label{fig:wnoise}
\end{center}
\end{figure}

In Fig. \ref{fig:wnoise} we can see how the EEH associated might change if eavesdropping, denoted by $EVE$, or any other source of noise occurs. The two EEHs are different, and just by comparing their structure we can note that something happened and the second protocol will result in a possible failure. Moreover, if an hypergraph is represented by one of the forbidden configurations, we might also infer that something within our specification, or the underlying language/hardware, is faulty.

\subsection{Technical Improvements and Implementations}
In this Section we presented two examples of EEHs relative to the teleportation protocol and a simplified version of a QKD protocol with $W$ states. 
We are currently developing an automatic tool that allows to build the EEH from its specification in a quantum programming language, e.g., Quipper. The tool, which is still in its infancies and by now requires a numerical representation of the states, is structured as follows: first, given a pure state in the standard computational basis, it returns its entangled graph by means of GSD and pairwise concurrence quantification/tangle. Since a quantum protocol can be thought as a sequence \emph{state} $\longrightarrow$ \emph{state}, the complete EEH is built after the computation of all the intermediate states. Since its behavior in time is represented as a causal process, we are investigating whether to use a guarded command language such as PRISM \cite{prism} to model the abstract EEH (before the generation of its explicit visual representation) and its transition from a state to the following one. 

\begin{figure}[h]
   \begin{minipage}{0.48\textwidth}
     \centering
     \includegraphics[width=0.5\textwidth]{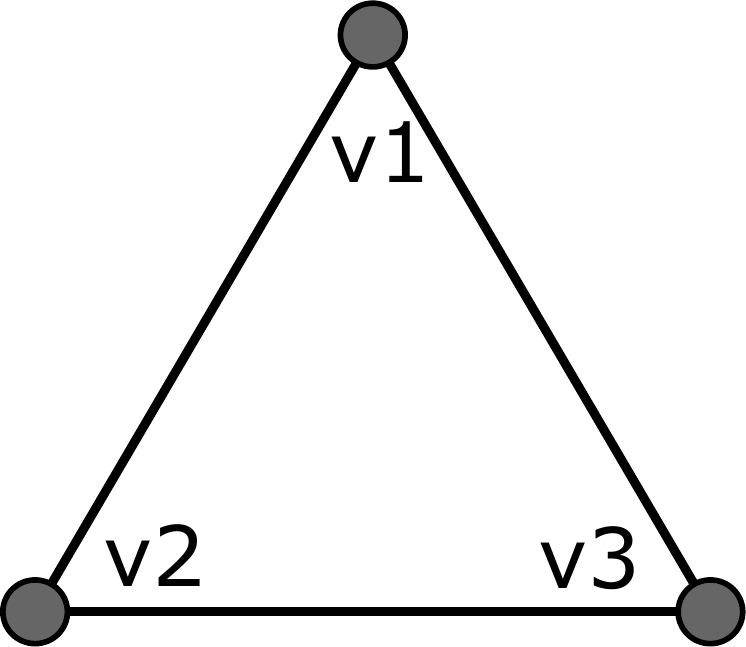}
   \end{minipage}\hfill
   \begin {minipage}{0.48\textwidth}
  \centering
     \includegraphics[width=\textwidth]{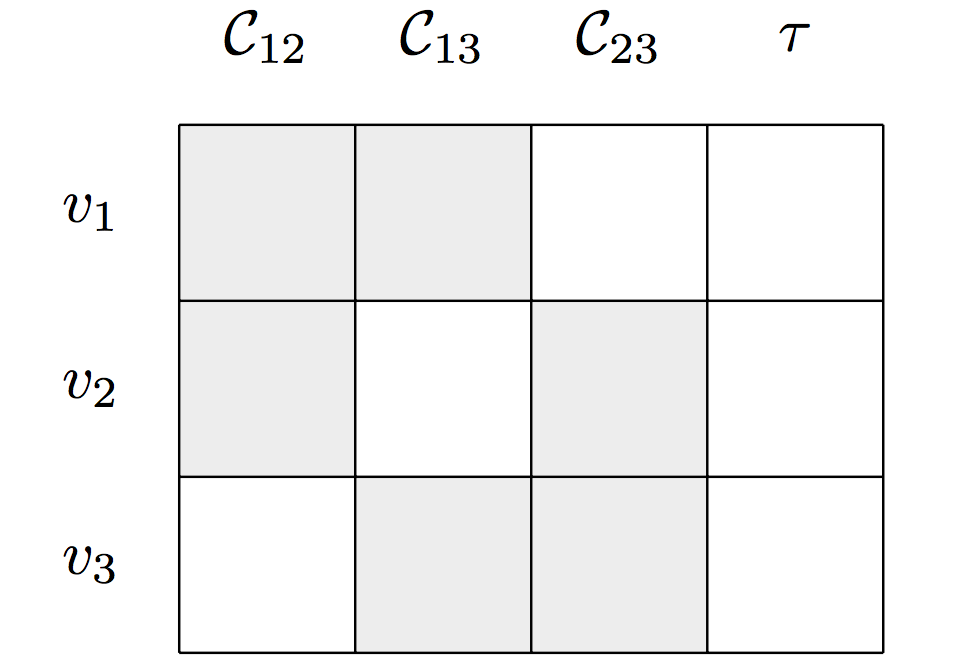}
   \end{minipage}
\protect\caption{Graphical representation of a $W$ state EH via incidence matrix}
\label{fig:grid}
\end{figure}

We are working on a visualization of the graph, which at each step will be represented as a lattice, or a grid of pixels. At present, we use as \emph{incidence matrix} of the hypergraph an instance of Table \ref{table:1}, where,  in order to render graphically the most explicit visual representation of the EEH topology, the columns represent concurrence and tangle and the rows represent the qubits. The changing in time of the shapes within the grid might be useful to investigate possible patterns associated to certain protocols and entanglement classes and the color can be related to the ``amount" of entanglement measured by concurrence and tangle. As an example, let us consider a $W$ state, whose entangled graph is taken unweighted for simplicity; by computing both pairwise concurrencies and tangle we obtain its incidence matrix:
$$M=\begin{pmatrix}
1 \ \ 1 \ \ 0 \ \ 0 \\ 1 \ \ 0 \ \ 1 \ \ 0 \\ 0 \ \ 1 \ \ 1 \ \ 0 
\end{pmatrix}$$ where $M_{i,j}=1$ means that the qubit $i$ has the property $j$. In our example, the qubit couples ($v_1,v_2$), ($v_1,v_3$), and ($v_2,v_3$) are pairwise entangled, but there is no tangle between them. This can be translated in the visual representation of Fig. \ref{fig:grid}.

\noindent More optimized structures, e.g., triangular graphs, could be proposed in an extended work.

We are also planning to provide the possibility to perform formal verification on the model obtained. Such verification can be done both temporally, i.e., by using model checking and a suitable temporal logic, e.g.,  QCTL, and spatially, by using a quantum extension of the SSTL spatial-temporal logic, presented in Ref. \cite{sstl}. The last one in particular is currently used to analyze emergent behaviors in dynamic systems (e.g., morphogenesis and pattern formation). A possible example of property to be verified on the grid can be the following (proper syntax and semantics are still to be investigated):
\texttt{Q=?[F(S[w(vi,vj)>0])]}.

\noindent where: 
\begin{itemize} 
\item $w(v_1,\dots,v_n)$ represents the weight (i.e., in this case the entanglement property) on a list of vertices;
\item  $F\phi$ is the temporal \emph{in the future} operator (as in the usual model checking theory) stating that, at a certain point during the execution of the protocol, the formula $\phi$ must be true;
\item  $S\phi$  is the spatial \emph{somewhere} operator,  which requires the formula $\phi$ to hold in a location reachable from the current one;
\item $Q\sim\mathcal{E}[\phi]$, with $\sim \in \{\geq,=,\leq\}$ is the QCTL quantum probability operator, which states that the probability that the formula $\phi$ is verified is bounded by a probability expressed by the quantum operation $\mathcal{E}$.
\end{itemize}
Roughly speaking, the formula investigates which is the probability that, starting at time $t_0$, in the future there will be an entangled couple in the graph. Again, we stress that this can be a direction in which this work can be extended.

The hypergraph structure is also suitable be used in machine learning tasks, both from the classification and the verification point of views, thus further investigations on a mixed approach (ML and MC techniques) should be carried on. Moreover, proofs about the mapping from the concrete domain (the quantum system) and the abstract one (the EEH) will be provided in the extended version of this work.

\section{Conclusions and Future Work}\label{sec: conclusion}

In this \emph{work-in-progress} paper we propose a new data structure, called Evolving Entangled Hypergraph (EEH), to represent protocols in which entanglement is an important property. The data structure is based on a classification of tripartite entanglement which, in turn, uses the concept of entangled hypergraph (EH). 

The EEH model is an abstract structure taking as input a class $L_i$ from a finite set $L$ of equivalence classes, evolving according some CPTP $\mathcal{E}$ and giving as output another class from the same set. 
This model allows us to track  interactions between qubits of a multipartite system and the evolution of the quantum system under consideration. It also allows to verify whether the entanglement properties are preserved by the time evolution or not. Since we want to model real quantum systems, we assume that the evolution may not be unitary and that some coupling with the environment may cause noise and decoherence, thus destroying the entanglement which is supposed to be preserved for the success of the quantum protocol under consideration.

The advantages in using the entanglement classification with the concept of EH is that it is a finitary method and that it is not required to compute the mathematical representation of operators. The hypergraph is computed by an algorithmic procedure, for which we are planning to build a tool to automate the process. In order to extract information about the entanglement from the system, modeled as an EEH, we can work on an high level structure. Then, by having an explicit classification declaring which are the allowed and forbidden configurations, we can verify whether the structure matches the requirements.

This method, which is still in its infancy, needs further investigation in different directions in order to provide both a reliable classification of multipartite entanglement for systems with more than three particles and a formal model, suitable to represent and formally verify quantum protocols. 
Moreover, it is required to understand whether the evolution of an EH does map it into a single class or into a set of classes. We also may need to deal with protocols which not use tripartite entanglement only but also the multipartite ones too, which now has been proposed as a resource for QKD protocols, for example in Ref. \cite{EKM}. 

Finally, we suggest that EEHs can be used to perform formal verification of quantum protocols. In particular, in the context of model-checking  EEHs provide a good model of the protocol execution, on which automatic and logical verification can be performed. Since they encode also a spatial dimension, we will be able to both define spatial locations of qubits (i.e., the vertices of the EH can be interpreted as qubit positions) and the temporal behavior of the system. We suggest that a spatial-temporal logic such as SSTL \cite{sstl} can be used to verify a protocol modelled by an EEH. In this way we shall be able to ask from the model whether a property of the system (e.g., entanglement) is preserved at a certain time in a specific location.

An important improvement that should also be investigated is whether EEHs can be used to model quantum systems of identical particles, by exploiting the spatial locations of qubits instead labeling them.

We are planning to extend the work proposed here with further formal details and proofs from both the classification (i.e., whether it is possible to use the presented method to classify also $\geq 4$-partite entanglement or not) and the abstract interpretation point of view. A tool which, given a state, provides the corresponding entangled hypergraph is currently under development.

\section*{Acknowledgment}
The first idea of this work formed at the Abdus Salam ICTP and the authors would like to acknowledge of this opportunity.


\newpage

\end{document}